%% file: main.tex
\documentclass[a4paper]{jpconf}
\usepackage{graphicx}
\input{Commands.tex}

\usepackage{wrapfig,lipsum}
\usepackage{subfig}
\usepackage{hyperref}

\begin{document}
\title{Sensitivity of the NEXT-100 detector to neutrinoless double beta
decay}

\author{N. L\'opez-March for the NEXT Collaboration}

\address{Instituto de F\'isica Corpuscular (IFIC), CSIC \& Universitat de Val\`encia \\Calle Catedr\'atico Jos\'e Beltr\'an, 2, 46980 Paterna, Valencia, Spain}

\ead{neuslopezmarch@ific.uv.es}

\begin{abstract}
A high pressure xenon gas time projection chamber with
electroluminescent amplification (EL HPGXe TPC) searching for the
neutrinoless double beta (\onu) decay offers: {\bf excellent energy resolution} \cite{Alvarez:2012kua,Lorca:2014sra} ($0.5-0.7\%$ FWHM 
at the $Q_{\beta\beta}$), by amplifying the ionization signal with electroluminescent light,
 {\bf and tracking capabilities} \cite{Ferrario:2015ina}, as demonstrated by the NEXT
 collaboration using two kg-scale prototypes. The NEXT collaboration is building an EL HPGXe TPC capable of holding
100 kg (NEXT-100) of xenon isotopically enriched in \Xe. The installation
and commissioning of the NEXT-100 detector at the Laboratorio
Subterr\'aneo de Canfranc (LSC) is planned for 2018. The current
estimated background level for the NEXT-100 detector is of
$4\times10^{-4}$ counts/keV-kg-yr or
 less in the energy region of interest \cite{Martin-Albo:2015rhw}. Assuming an energy resolution of 0.75\%
 FWHM at the $Q_{\beta\beta}$ and a \onu~signal efficiency of about
 28\%, this gives an expected sensitivity (at 90\% CL) to the
 \onu~decay half life of $T^{0\nu}_{1/2}>6.0\times10^{25}$ yr for an
 exposure of 275 kg~yr. A first phase of the NEXT experiment, called
 NEW, is currently being commissioned at the LSC. The NEW detector is a scale 1:2 in size
(1:10 in mass) of the NEXT-100 detector using the same materials and
photosensors and will be used to perform a characterization of the
\onu~backgrounds and a measurement of the standard double beta decay
with neutrinos (\twonu). An 8 sigma significance
for the \twonu~signal in the NEW detector has been estimated for a 100-day run.
\end{abstract}

\section{Introduction}
\vspace{0.2cm}

One of the most promising technologies to search for the \onu~decay is 
an asymmetric high pressure xenon gas (HPGXe) time projection chamber (TPC) with 
electroluminescent (EL) amplification. The
NEXT collaboration is building an EL HPGXe TPC capable of holding
100 kg (NEXT-100) of xenon isotopically enriched with \Xe. The installation of NEXT-100 at the LSC is planned for 2018. This
technology offers {\bf excellent energy resolution} \cite{Alvarez:2012kua,Lorca:2014sra} ($0.5-0.7\%$ FWHM 
at the $Q_{\beta\beta}$), by amplifying the ionization signal with electroluminescent light,
 {\bf and tracking capabilities} \cite{Ferrario:2015ina}, as demonstrated by the NEXT
 collaboration using two kg-scale prototypes.

The EL amplification is essential to get a linear gain avoiding
avalanche fluctuations and to fully exploit the excellent Fano factor
of xenon in gas to obtain excellent energy resolution. In NEXT, the EL light
is collected by an array of photomultipliers located behind the
cathode (energy measurement) as well as by a dense array of silicon
photomultipliers (topology measurement) located behind the anode. The tracking capability allows for the distinction of signal events (the two electrons
emitted in a \onu~decay), reconstructed as a
continuous track with larger energy depositions (blobs) at both ends, and
background events (mainly due to single electrons, from ${}^{208}$Tl and
${}^{214}$Bi, with kinetic energy comparable to the end-point of the \onu~decay) reconstructed as a
track with only one end-of-track blob \cite{paola}.

NEXT is an international collaboration that includes research groups
from Spain, Portugal, USA, Russia, and Colombia. 
The NEXT research program has been organized into four stages: 1)
demonstration of the EL HPGXe TPC technology with $\sim$ 1 kg detectors
(NEXT-DEMO and NEXT-DBDM); 2) characterization of the backgrounds for the \onu~signal and
measurement of the \twonu~decay with a 10 kg detector called
NEXT-WHITE (NEW) at the Laboratorio Subterr\'aneo de Canfranc (LSC); 
3) search for the \onu~decay with the NEXT-100 detector at the LSC and 4) scale
up and further development to reduce backgrounds and enhance the
topological signature for a 1 tonne-scale EL HPGXe TPC.

\section{Sensitivity of the NEXT-100 detector to \onu~decay}
\vspace{0.2cm}

The most important background source in NEXT comes from radioactive
impurities in the detector components from the uranium and thorium series, 
particularly ${}^{208}$Tl and ${}^{214}$Bi, whose photo-peaks lie around the
hypothetical \onu~peak of \Xe~(Q$_{\beta\beta}$ = 2.458 MeV).

A thorough campaign of radiopurity measurements have been performed from 2011 to
2016 to estimate the activity of the ${}^{208}$Tl and ${}^{214}$Bi
background sources in the most relevant components of the NEXT-100
detector. On the other hand, an estimate of the \onu~signal and background detection
efficiencies, for the \onu~event selection in the NEXT-100 detector, has
been evaluated using Monte Carlo (MC) simulations. 
\begin{wrapfigure}{r}{6.5cm}
\includegraphics[scale=0.35]{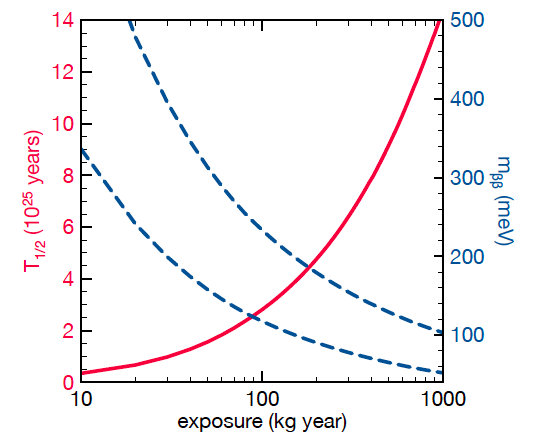}
\caption{Sensitivity (at 90 \% C.L) of the NEXT-100 detector to the
  \onu~half-life and to the \mbb~as a function of the accumulated exposure for an estimated
  background rate of $4\times10^{-4}$ counts/keV-kg-yr in the ROI.}
\label{next100}
\end{wrapfigure}
A \onu~candidate event requires that: 1) only one track is
reconstructed fully contained within the fiducial volume of the
detector (defined by excluding a region of 2 cm around the boundaries
of the active volume); 2) the reconstructed track features a blob at both ends; 3) the energy of the event is within the region of interest (ROI)
2.448 $<$ E $<$ 2.477 MeV. This selection gives an efficiency of 28\%
for \onu~signal events, while the natural radioactive
backgrounds, ${}^{208}$Tl and ${}^{214}$Bi, are suppressed by more
than 6 orders of magnitude and the background from \twonu~decays is
completely negligible. 

Taking into account the contribution of each detector subsystem from
the material-screening measurements and the \onu~event selection, the
estimated overall background rate in NEXT-100 is of
$4\times10^{-4}$ counts/keV-kg-yr or less in the ROI \cite{Martin-Albo:2015rhw}. Assuming an energy
resolution of 0.75\% FWHM at the $Q_{\beta\beta}$ and a \onu~signal efficiency of about 28\%, this
gives an expected sensitivity (at 90\% C.L) to the \onu~decay
half life of $T^{0\nu}_{1/2}>6.0\times10^{25}$ yr for an exposure of
275 kg~yr. Figure \ref{next100} shows the expected sensitivity (at 90\% C.L) of
the NEXT-100 detector to the \onu~half-life and the corresponding
sensitivity to the neutrino Majorana mass, \mbb, as a function of
accumulated exposure for the largest and smallest nuclear matrix
element estimates (blue dashed curves).

\section{NEXT-WHITE background expectations and sensitivity to \twonu~decay}
\vspace{0.2cm}

The NEW detector is the first phase of the NEXT experiment to operate
underground and is currently being commissioned at the LSC. NEW is a scale 1:2 in size
(1:10 in mass) of NEXT-100 using the same materials and
photosensors \cite{miquel}. A similar background model as for the NEXT-100 detector
has been developed for the NEW detector considering both depleted and
enriched \Xe~runs.
The NEW background model is particularly detailed, as it considers
potential contributions from four isotopes (${}^{40}$K,  ${}^{60}$Co ,
${}^{214}$Bi, ${}^{208}$Tl) and 17 detector components, for a total of 68
background sources. 

\begin{figure}[htb]
 \begin{center}
 \includegraphics [scale=0.35]{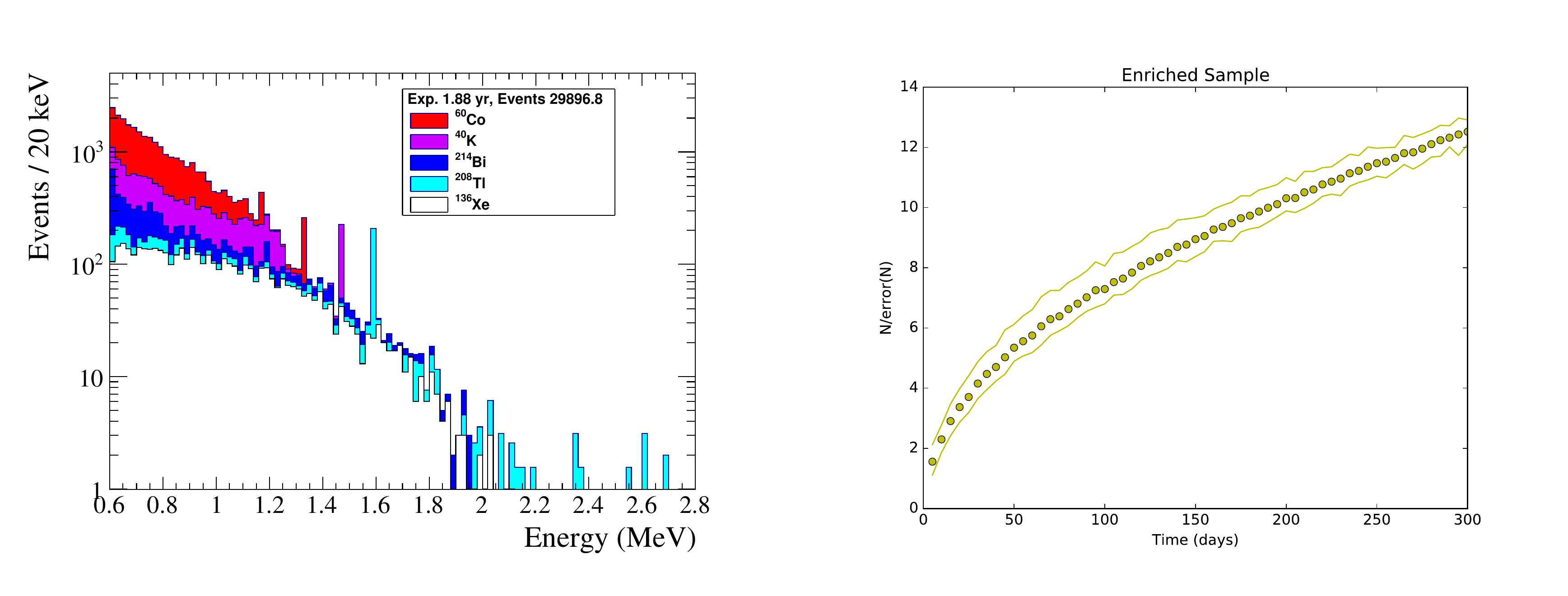} %
   \caption{Left: expected energy spectrum in NEW for the
     \twonu~selection. Right: contributions to the background rate of NEXT-100 from different detector components. An asterisk (*) indicates that the contribution corresponds to a positive measurement of the activity of the material. }
    \label{NEW}
  \end{center}
\end{figure}

A similar event reconstruction and selection as the one developed for
the NEXT-100 \onu~analysis has been performed using MC simulations to search for \twonu~in
NEW, the main difference being the looser energy requirement (0.6 $<$
E $<$ 2.8 MeV). Figure  \ref{NEW}-left shows the expected energy spectrum
for events passing the \twonu~event selection. Taking into account the
material-screening measurements and the \twonu~event selection, a
significance of $~$8 sigmas for a 100-day run is estimated for the
\twonu~measurement, as shown in Fig. \ref{NEW}-right.

\section{Prospects}
Further developments to reduce backgrounds and to fully exploit the
potential of the tracking signature (e.g. using deep learning techniques
and different gas mixtures to reduce diffusion) are being studied by the
NEXT collaboration to enhance the sensitivity to the \onu~decay for a one tonne-scale EL HPGXe TPC.

\ack
The NEXT Collaboration acknowledges support from the following agencies and institutions:
the European Research Council (ERC) under the Advanced Grant 339787-NEXT;
the Ministerio de Econom\'ia y Competitividad of Spain and FEDER under grants CONSOLIDER-Ingenio
2010 CSD2008-0037 (CUP), FIS2014-53371-C04 and the Severo Ochoa Program
SEV-2014-0398; GVA under grant PROMETEO/2016/120. Fermilab is operated by Fermi Research Alliance, LLC under Contract No. DE-AC02-07CH11359 with the United States Department of Energy.

\bibliographystyle{iopart-num}

\section*{References}

\bibliography{refs}

\end{document}

%% file: Commands.tex
\newcommand{\onu}{\ensuremath{0\nu\beta\beta}}
\newcommand{\twonu}{\ensuremath{2\nu\beta\beta}}

\newcommand{\mbb}{\ensuremath{m_{\beta\beta}}}

\newcommand{\Xe}{\ensuremath{{}^{136}\rm Xe}~}
